\documentclass[aps,pra,twocolumn,showpacs,amsmath,amssymb,superscriptaddress,superscriptaddress]{revtex4-1}

\usepackage{graphicx}
\usepackage{epstopdf}
\usepackage{bm}
\usepackage{times}
\usepackage{comment}
\usepackage{dsfont}
\usepackage{color}
\definecolor{blue}{RGB}{0,0,255}
\usepackage[colorlinks,citecolor=blue,linkcolor=blue,urlcolor=blue]{hyperref}

\begin{document}
\def\BE{\begin{equation}}
\def\EE{\end{equation}}
\def\BY{\begin{eqnarray}}
\def\EY{\end{eqnarray}}
\def\BI{\begin{itemize}}
\def\EI{\end{itemize}}
\def\L{\label}
\def\nn{\nonumber}
\def\({\left (}
\def\){\right)}
\def\[{\left [}
\def\]{\right]}
\def\<{\langle}
\def\>{\rangle}
\def\o{\overline}
\def\BA{\begin{array}}
\def\EA{\end{array}}
\def\dsp{\displaystyle}
\def\ds{\displaystyle}
\def\k{\kappa}
\def\dd{\delta}
\def\D{\Delta}
\def\w{\omega}
\def\W{\Omega}
\def\v{\nu}
\def\a{\alpha}
\def\b{\beta}
\def\e{\varepsilon}
\def\ee{\text{e}}
\def\d{\partial}
\def\g{\gamma}
\def\G{\Gamma}
\def\tt{\theta}
\def\t{\tau}
\def\s{\sigma}
\def\+{\dag}
\def\8{\infty}
\def\x{\xi}
\def\m{\mu}
\def\l{\lambda}
\def\ii{\text{i}}
\def\={\approx}
\def\xc{\frac{2x}{c}}
\def\->{\rightarrow}
\def\r{\vec{r}}
\def\k{\vec{k}}
\def\sinc{\mathrm{sinc}}
\def\xx{\textbf{x}}
\def\yy{\textbf{y}}
\def\qq{\textbf{q}}
\def\rr{\boldsymbol{\rho}}
\newcommand{\ud}{\,\mathrm{d}} 
\def\out{|\textrm{out}\rangle}
\def\inn{|\textrm{in}\rangle}
\def\sqz{|\textrm{sqz}\rangle}
\def\A{\text{A}}
\def\B{\text{B}}
\def\AB{\text{AB}}
\def\where{\text{where:}}
\def\F{{\cal F}}
\def\A{{\cal A}}
\def\in{\text{in}}

\title{Heralded temporal shaping of single photons enabled by entanglement}

\author{Valentin Averchenko}
\email{valentin.averchenko@gmail.com}
\affiliation{Max-Planck-Institut f\"ur die Physik des Lichts, G\"{u}nther-Scharowsky-Stra{\ss}e 1, Bau 24, 91058 Erlangen, Germany}

\author{Denis Sych}
\affiliation{Max-Planck-Institut f\"ur die Physik des Lichts, G\"{u}nther-Scharowsky-Stra{\ss}e 1, Bau 24, 91058 Erlangen, Germany}

\author{Gerd Leuchs}
\affiliation{Max-Planck-Institut f\"ur die Physik des Lichts, G\"{u}nther-Scharowsky-Stra{\ss}e 1, Bau 24, 91058 Erlangen, Germany}
\affiliation{Department f\"{u}r Physik, Universit\"{a}t Erlangen-N\"{u}rnberg, Staudtstra{\ss}e 7, Bau 2, 91058 Erlangen, Germany} 

\date{\today}

\begin{abstract}

We propose a method to produce pure single photons with an arbitrary designed temporal shape in a heralded, lossless and scalable way.
As the indispensable resource, the method uses pairs of time-energy entangled photons.
To accomplish the shaping, one photon of a pair undergoes temporal modulation according to the desired shape. 
Subsequent frequency-resolving detection of the photon heralds its entangled counterpart 
in a pure quantum state with a temporal shape non-locally affected by the modulation.
We found conditions for the shape of the heralded photon to reproduce the modulation function.
The method can be implemented with various sources of time-energy entangled photons. In particular, using entangled photons from the parametric down-conversion the method enables generation of pure photons with tunable shape within unprecedentedly broad range of temporal durations - from tenths of femtoseconds to microseconds.
Proposed shaping of single photons will push forward implementation of scalable multidimensional quantum information protocols, efficient photon-matter coupling and quantum control at the level of single quanta.

\end{abstract}


\maketitle

Single photons are indispensable tool in quantum information, quantum communication and quantum metrology \cite{Eisaman2011}. Depending on specific application, photons must be tuned in wavelength, bandwidth, polarization, and other degrees of freedom.
For example, to transmit information over optical fibers one uses photons at telecom wavelengths, while photon-atom coupling requires wavelength and bandwidth of photons matched to the corresponding atomic transition.

It has been realized in the last decade that full control over the spatio-temporal shape of a single photon light  \cite{Raymer2012, Walmsley2015} is required in a number of applications.
For example, photons, having identical Lorentzian spectral shapes, can exhibit opposite temporal shapes  - exponential decaying or rising - depending on spectral distribution of the phase.
As a result, photons with exponential decaying shape excite a two-level atom in free space only with 50$\%$ efficiency, while exponentially rising photons with 100$\%$ efficiency  \cite{Stobinska2009, Wang2011, Aljunid2013}.
There are other situations where the temporal photon shape is important: symmetric shape is optimal for cavity QED quantum communication \cite{Cirac1997}; Gaussian shape is superior for experiments relying on single-photon interference \cite{Rohde2005}. Furthermore, development of sources of shaped photons will push forward photonic implementation of high-dimensional quantum information protocols, e.g. quantum key distribution \cite{Bechmann:00,Bourennane:01,Cerf:02,Sych:04,Sych:05, Brecht2015}.

Present-day methods for producing temporally shaped photons can be divided in two groups: single photons are shaped either during the generation process or photon shaping is performed after the generation.
Methods of the first group are typically based on control of quantum emitters (single atom \cite{Kuhn2002, McKeever2004}, ion \cite{Keller2004a}, atomic ensemble \cite{Eisaman2004}, quantum dot \cite{Matthiesen2013}). Development of these methods is promising for deterministic production of shaped photons, however the methods often require sophisticated and costly setups to manipulate single quantum emitters. 
Also wavelengths of generated photons are limited to resonant transitions of quantum emitters that restricts tunability of these methods.
Methods of the second group are based on direct spectral filtering \cite{Baek2008} or amplitude-phase modulation \cite{Kolchin2008, Specht2009} of single photons. The methods require less elaborated setups then methods of the first group and show higher flexibility in wavelengths and durations of produced photons.
However losses inherent to the filtering and modulation of single photon light lead to probabilistic production of shaped photons, which prevents efficient production of multiple of shaped photons and hinders realization of scalable photonic networks.

	\begin{figure}
	\center{\includegraphics[width=0.9\linewidth]{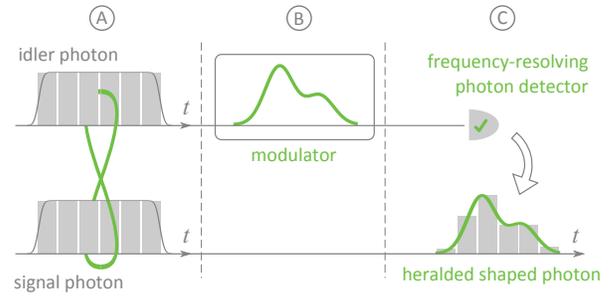}}
	\caption{Schematic depiction of the temporal shaping method: A) To produce shaped single photon we start with a pair of time-energy entangled photons, conventionally called signal and idler. 
B) The idler photon passes through a temporal modulator with the desired transmission function. C) After the modulator, the idler photon is detected by a frequency-resolving single-photon detector. Provided the successful detection of the idler photon (``click'' event), the conditional heralded shape of the signal photon (shown by the green solid line) follows the modulator function (see Eqs.~(\ref{psi-t-w}) and (\ref{psi-t-w-1})). }
	\label{scheme}
	\end{figure}

In this work we show a method to produce single photons with any desired temporal shape in a pure quantum state. 
Shaped photons are produced in a heralded, conditionally lossless and scalable way.
The method is based on general photon shaping idea proposed in \cite{Sych2016}. 
Furthermore the method can be viewed as remote preparation of time-encoded single-photon state \cite{Zavatta2006} or as "ghost" interference \cite{Strekalov:95} in the time domain \cite{Bellini2003}.

Principal scheme of the method is presented in Fig.~\ref{scheme}. As a resource, the method uses pairs of time-energy entangled photons \cite{Franson1989} that we conventionally call signal and idler. 
Required entangled photons can be produced via different optical nonlinear processes, particularly, via optical parametric down-conversion.
To produce shaped signal photon we perform temporal modulation of idler photon according to the desired shape and subsequent frequency-resolving detection of the photon.
The frequency-resolving detection measures the frequency of the idler photon in contrast to the time-resolving detection that measures the photon arrival time. We show further that under these conditions and due to the initial entanglement, successful detection of the idler photon heralds pure signal photon with the temporal shape affected by the modulation in the idler arm. 

We stress, that a potentially lossy shaper (temporal modulator) is brought into the idler arm and only successful measurement outcomes herald shaped signal photons \cite{Sych2016}.
Due to that, the shaping method is conditionally lossless, since the detector ``click'' heralds a shaped signal photon with certainty (in the absence of technical imperfections in the idler arm). This feature enables to produce multiple of shaped photons in a scalable way, in contrast to traditional method of direct modulation of the heralded photons.

Now we present formal description of the method outlined above and derive explicit expressions for the temporal shape of the heralded signal photon depending on: (A) the initial quantum state of the photon pair, (B) modulation applied to the idler photon, and (C) outcome of the frequency-resolving measurement of the idler photon (see Fig.~\ref{scheme}).

We start with the main resource of the method, namely, signal and idler photons in an entangled state. We assume 
that the entangled state is described by a joint probability amplitude $\Psi(t,t')$ which has a simple physical meaning~--- its modulus squared gives joint probability density to detect signal photon at a time instant $t$ and idler photon at a time instant $t'$ (here and below arguments with prime refer to the idler photon and ones without prime~--- to the signal photon, unless otherwise stated). One says the photons are entangled when the function $\Psi(t,t')$ can not be factorized with respect to $t$ and $t'$.

Now we show that applying the temporal modulation to the idler photon and performing its subsequent frequency-resolving measurement one can herald pure signal photon with the temporal shape non-locally affected by the modulator.

Temporal modulator placed in the idler arm can be considered as time-dependent transmission of the optical filed and described by a complex function $\A(t)$, such that $|\A(t)|\leq 1$. We assume that transmission is nonzero only within a finite time window $t_m$.
If the idler photon passed the modulator then the joint amplitude of the photon pair transforms as 
	\begin{equation}
	\Psi(t,t') \rightarrow \A(t') \Psi (t, t').
	\L{mod}
	\end{equation}

After the modulator, we perform frequency-resolving measurement of the idler photon. Detection of the idler photon with a frequency $\w'$ heralds signal photon in a conditional quantum state, described by the corresponding probability amplitude $\psi(t|\w')$. 
Here and further frequencies of photons are counted with respect to carrying ones.
The better resolution of the photon frequency measurement the more uncertain time instant when the photon has passed the modulator according to the time-energy uncertainty relation. 
Ideal frequency-resolving measurement of the idler photon ensures that one can not principally determine at which time instant the photon has passed the modulator, therefore resulting probability amplitude is a coherent sum of amplitudes representing all possibilities 
\footnote{The resulting probability amplitude can be also obtained in the formalism of bra-ket vectors as the projection of a two-photon state on an idler single-photon state with a well-defined energy, i.e.:
{$|\psi\rangle_s\propto {}_i\langle \omega'|\Psi\rangle_{si}$}, where {$|\Psi\rangle_{si} = \iint\,\mathrm{d} t \,\mathrm{d} t' \Psi(t,t') |t\rangle_s|t'\rangle_i$} and {$|\omega'\rangle_{i} = \int \,\mathrm{d} t' e^{-\text{i}\omega' t'}|t'\rangle_i/\sqrt{2\pi}$}}
and up to normalization it can be written as
	\begin{equation}
	\psi(t|\w') \propto \int \ud t' \A(t') \Psi (t, t') e^{\i \w' t'}. 
	\L{psi-t-w}
	\end{equation}
Obtained probability amplitude has the following meaning - its modulus squared defines probability density distribution to detect the signal photon at a time instant $t$ given the idler photon measurement outcome $\w'$. We refer this amplitude (\ref{psi-t-w}) as conditional or heralded temporal shape of the signal photon. Proportionality symbol denotes that the amplitude is not normalized, which is used further just for convenience.

The generic expression for the shape of the signal photon (\ref{psi-t-w}) depends on the initial joint state $\Psi(t,t')$, applied modulation ${\cal A}(t)$ and the obtained measurement outcome $\w'$ in the idler arm. The joint control of these parameters tailors the photon shape. 

	\begin{figure}
	\center{\includegraphics[width=0.75\linewidth]{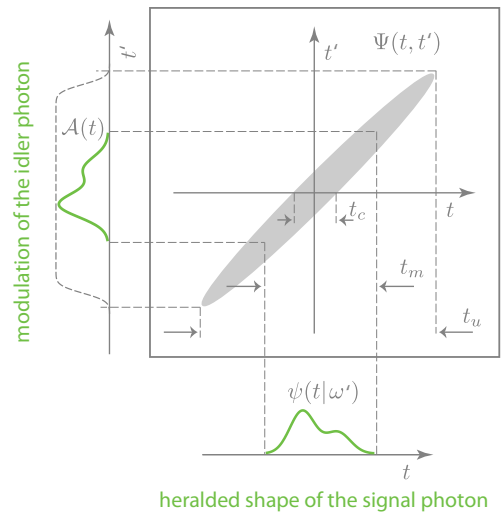}}
	\caption{Influence of the modulator, applied to the idler photon, on the heralded shape of the signal photon. Due to the temporal correlations between the signal and idler photons (the joint probability density $\Psi(t,t^\prime)$ is visualized by the ellipse), and frequency-resolving detection of the idler photon, the shape of the signal photon follows the modulator transmission function $\A(t)$. The condition  $t_c\ll t_m < t_u$
corresponds to the case of perfect correlations, i.e. the joint amplitude is approximated as $\Psi(t,t^\prime)\simeq\delta(t-t^\prime)$ and the probability density can be schematically depicted as infinitely elongated and narrow ellipse.
}
	\label{g_tt}
	\end{figure}

Now we analyze the shape of the heralded photon depending on joint state $\Psi(t,t')$ of photon pairs.
We characterize the joint state with two time scales: unconditional $t_u$  and conditional $t_c$  temporal widths of photons distributions (see Fig. (\ref{g_tt})) that have the following meaning. Unconditional individual measurements of the signal and idler photons with fast time-resolving single-photon ``click'' detectors show statistical uncertainty with a characteristic time $t_u$. Measurements of the signal photon, conditioned on the measurement of the idler photon at a particular time instant (or vice versa, measurements of the idler photon conditioned on a particular detection time of the signal photon), indicate statistical spread $t_c$ which is smaller than $t_u$ due to the temporal correlations between the signal and idler photons. The stronger the temporal correlations between the photons, the higher the ratio $t_u/t_c$ (see, for example, \cite{Just2013}).
Here and further all time scales are larger than an optical cycle of light fields.

Consider the physical situation, when the signal and idler photons are entangled such that their correlation time $t_c$ is the shortest one, their unconditional temporal width $t_u$ is much longer, and the modulation time $t_m$ is in between:
	\begin{equation}
	t_c \ll t_m < t_u.
	\label{cond-1}
	\end{equation}
The joint state can be approximated as maximally entangled $\Psi(t,t') \propto \dd(t-t')$ in the integral (\ref{psi-t-w}), hence the temporal shape of the heralded signal photon is
	\begin{align}
	\psi(t|\w') \propto \A(t) \; e^{\i \w' t}. 
	\L{psi-t-w-1}
	\end{align}
The expression represents the main result of the paper and shows that temporal shape of the heralded photon is defined by the modulation function in the heralding arm.

We stress that to herald pure shaped signal photon the following conditions are crucial: initial entanglement between signal and idler photons, modulation of the idler photon, frequency-resolving measurement of the idler photon. We consider these conditions in more details.
First, if photon pairs posses only classical correlations and described by a mixed quantum state, then implementation of the proposed method results in signal photons, heralded in a mixed state \cite{Ryczkowski2016, Sych2016}. 
Second, in the extreme case of disentangled and uncorrelated photon pairs detection of the idler photon after the modulator does not affect in the post-selection temporal shape of the signal photon and its conditional state does not change. Third, if instead of the frequency-resolved measurement we use the time-resolved one then detection of the idler photon after the modulator at time instant $t'$ heralds the signal photon in the conditional state not affected by the modulation. It can be seen from (\ref{mod}), where left and right hand sides are the same functions of $t$.

Now we analyze implementation of frequency-resolving measurement of the idler photon required in the proposed method. The measurement can be realized with a spectral filter followed by a single-photon time-resolving ''click'' detector. The filter can be implemented with an optical cavity, atomic ensemble, or another frequency-selective element that transmits idler photons around a particular frequency that we denote $\w'$.
The frequency is counted with respect to the carrying frequency of the photons. 
We analyze this part of the shaping method in more details using a specific model of the filter. 
We describe the filter operation in the time domain.
Filter delays the idler photon by the time $\t$ with the probability proportional to the modulus squared of impulse response of the filter, that we denote as follows: ${\cal F}(\t) e^{-\ii\w'\t}$. 
Due to the probabilistic delay, the filter ``removes information'' at which time instant the photon has passed the modulator. 
Click of the detector at a time instant $t'$ heralds the signal photon with the conditional amplitude given by the convolution of the filter response and joint amplitude of photons before the filter (\ref{mod}): 
	\begin{align}
	\psi(t|t',\w') \propto \int \ud \t \F(t'-\t) e^{-\ii\w'(t'-\t)} \A(\t) \Psi (t, \t)
	\end{align}
Let us consider the case of perfectly correlated photons satisfying the condition (\ref{cond-1}). Then the shape of the heralded photon is given by the expression: $\psi(t|t',\w')~\propto~\A(t)~\F(t'~-~t)~e^{-\ii\w'(t'-t)}$. The shape is defined by the modulation function $\A(t)$ multiplied with the impulse response of the filter $\F(t'-t) e^{-\ii\w'(t'-t)}$. 
Consider an experimental situation such that
characteristic response time of the filter, defined by inverse filter bandwidth $\w_f^{-1}$, is longer than the modulation time, i.e.:
	\begin{align}
	& t_m \ll \w_f^{-1}
	\L{tf}
	\end{align}
and one can put approximately $\F(t'-t)\approx\text{const}$. Then the shape of the heralded photon is defined by the modulation function: $\psi(t|t',\w') \propto \A(t) e^{\ii\w' t}$, provided post-selection on the idler photons detected after the end of the modulation.
It is the same result as (\ref{psi-t-w-1}) obtained for a particular model of frequency resolving measurement.
We conclude that the heralded photon is synchronized with the modulation and its shape does not depend on a specific  heralding instant $t'$ --- all idler clicks after the end of the modulation are suitable as heralds. Furthermore, due to the condition (\ref{tf}), most of the heralding clicks happen after the modulation.

Heralding rate of shaped signal photons in the proposed method can be estimated as
	\begin{align}
	R \approx \w_f t_c\frac{t_m}{t_u}.
	\L{psi-t}
	\end{align}
The expression has the following meaning~--- heralding rate is defined by the fraction of idler photons that pass both the temporal modulator $t_m/t_u$ and the spectral filter $\w_f t_c$, giving the bandwidths of photons and the filter are $t_c^{-1}$ and $\w_f$ respectively. The rate can be increased if instead of modulating the idler photon one modulates generation of the entangled pairs, i.e. produces pairs of photons in the pulsed way. 
For example, one can employ parametric down-conversion with the pulsed pump with the shaped envelope ${\cal A}(t)$ \cite{Ou1997, Grosshans2001, Aichele2002, Kalachev2010}. 
Then frequency resolved detection of the idler photon heralds the signal photon with the shape (\ref{psi-t-w-1}) defined by the pump envelope. Corresponding heralding rate is $R \approx \w_f t_c$ which is higher then (\ref{psi-t}).
We notice, that this approach provides not only a higher heralding rate of shaped photons, but also reaches femtosecond duration of the heralded shaped photons~--- this range is not easily achievable with the direct modulation of idler photons using acousto- or electro-optical modulators.

One may naturally expect, that experimental imperfections can affect the parameters of the shaped signal photons. Let us consider an example. 
Expression (\ref{psi-t-w-1}) shows that
carrying frequency of the heralded photon depends  on an outcome $\w'$ of the frequency resolving measurement.
Statistical uncertainty of the measurement, that we denote as $\w_d$, results in the corresponding uncertainty of the carrying frequency of heralded photons. The effect is negligible when the uncertainty is smaller then bandwidth of the heralded photon, i.e. $t_m \w_d\ll 1$. 
The effect can be quantitatively characterized calculating purity of the quantum state of heralded photon depending on the uncertainty $\w_d$. 
Let us model statistical distribution of measurement outcomes $\w'$ with the function $\g(\w'-\w_0')$, given the frequency of the impinging photon is $\w_0'$.
Then detector click at the frequency $\w_0'$ heralds the signal photon, whose state is described by the following density matrix
	\begin{align}
	& \rho(t,t'|\w_0') = \int\ud\w' \; \g(\w'-\w_0') \psi(t|\w') \psi^*(t'|\w')
	\L{rho-s}
	\end{align}
Here both arguments $t$ and $t'$ refer to the signal photon and $\psi(t|\w')$ is given by expression (\ref{psi-t-w}).
We calculate purity $\pi \equiv \text{Tr}\,\rho^2$ of this state as a function of four experimentally relevant parameters $t_c$, $t_u$, $t_m$, and $\w_d$ using the following Gaussian approximations
	\begin{align}
	& \Psi(t, t') = \ee^{-(t-t')^2/2 t_c^2} \; \ee^{-(t+t')^2/2 t_u^2}\sqrt{2/\pi t_c t_u},\\
	& \A(t) = \ee^{-t^2/2 t_m^2},\\
	& \g(\w) = \ee^{-\w^2/\w_d^2}/\sqrt{\pi} \w_d.
	\end{align}
Resulting general expression for the purity
\footnote{
	$
	\pi =\sqrt{	\frac{(t_c^2 + 4t_m^2 + t_u^2)(t_m^2 t_u^2 + t_c^2(t_u^2 + t_m^2(1+ t_u^2\w_d^2)))}{(t_u^2 t_m^2 + t_c^2(t_u^2+t_m^2))(t_c^2 + 4t_m^2 + t_u^2 + t_m^2\w_d^2(t_c^2+t_u^2))}}	
	$
} 
can be applied for various experimental conditions. In a particular case (\ref{cond-1}) of highly correlated photon pairs, such that $t_u \rightarrow \infty$ and $t_c \rightarrow 0$, we get the following estimation:
	\begin{align}
	& \pi \approx \frac{1}{\sqrt{1+ t_m^2 \w_d^2}}.
	\L{purity}
	\end{align}
The state of the heralded photon  is approximately pure when the uncertainty of the frequency-resolved  measurement is smaller than the modulation bandwidth, i.e. $t_m \w_d\ll 1$.

We estimate parameters of shaped single photons experimentally achievable with the proposed method.
The condition (\ref{cond-1}) shows that possible duration of shaped photons is limited by temporal characteristics of entangled photons produced by a photon source. 
For a given source the method provides to produce shaped photons longer then correlation time of photon pairs $t_c$ and shorter then their unconditional temporal spread  $t_u$. Furthermore, the condition (\ref{tf}) shows that maximal possible duration of heralded photon is also limited by the bandwidth of frequency-resolving measurement.
Up to date, the most widely used sources of entangled photons are based on
spontaneous parametric down-conversion in second-order non-linear medium that can be implemented both in single-pass \cite{Friberg1985} and cavity-assisted \cite{Ou1999} configurations. 
Parametric sources provide entangled photons in various  spectral ranges and bandwidths.
Unconditional temporal spread of photons $t_u$ is typically defined by properties of the strong coherent pump light that drives parametric process. In the continuous regime, it is the coherence time of the pump ($\mu$s-ms range). In the pulsed regime it is the duration of pump pulses (up to tens of fs).
Correlation time of parametrically generated photons $t_c$ is defined either by the phase-matching bandwidth in the non-linear media (typically, fs-ps range) or cavity bandwidth in a cavity-assisted configuration (ns range).
Therefore, the use of parametric down-conversion as a resource of time-energy entangled photons enables generation of shaped photons in  a very broad range of temporal durations - from fs to $\mu$s.

In conclusion, we have proposed a method to produce single photons with an arbitrary temporal shape in a pure quantum state. 
The method utilizes time-energy entangled photon pairs as the indispensable resource.
Shaped photons are produced in a heralded and lossless manner upon temporal modulation and  successful frequency-resolving measurement of their entangled counterparts. 
We derive the explicit expression for the temporal shape of heralded photons. The shape is determined by the initial entanglement of photon pairs, modulation function and measurement outcome in the heralding arm. The joint control of these parameters affects the shape of heralded photons.

Two features of the method are worth noting. First, shaped photons are produced in a conditionally lossless way. It enables scalable production of multiple of shaped photons. This feature is crucial for realization of scalable quantum photonic networks.

Second, the method is ready to implement with present-day photonic technologies, allowing to produce shaped photons in a broad range of wavelengths and bandwidths. 
In particular, using parametric down-conversion as a source of entangled photons, the method is promising to produce shaped photons in immense range of temporal durations - from tenths of femtoseconds to microseconds.
Experimental results on non-local temporal modulation \cite{Sensarn2009} and remote preparation of time-encoded single-photon ebits \cite{Zavatta2006} confirm experimental feasibility of the proposed shaping method.

To experimentally verify shape of heralded photon the method can be accompanied with one of tomographic techniques, provided informationally complete measurements performed on the photon \cite{Polycarpou2012, Morin2013, Qin2015, Sych:12,Bent:15}. 

We expect that simplicity and scalability of the proposed method will make generation of temporally shaped photons a common practice in quantum optics and will push forward experiments that require shaped photons, e.g. efficient photon-matter coupling, multidimensional quantum information protocols, coherent control of quantum systems at the level of single quanta, to name a few.
Besides, proposed shaping method reveals a novel practical application of the quantum entanglement. 

We acknowledge Alessandra Gatti, Paul Kwiat, Christoph Marquardt, Gerhard Schunk and Ulrich Vogel for useful discussions.
V.A. acknowledges funding by the Max Planck Society.

\bibliography{bibl}
\end{document}